\documentclass[journal]{vgtc}            

\onlineid{1830}

\preprinttext{To appear in IEEE Transactions on Visualization and Computer Graphics.}

\vgtccategory{Research}

\vgtcpapertype{system}

\newcommand{\myparagraph}[1]{\noindent{\textbf{#1}.}}
\newcommand{\systemname}[0]{Urbanite\xspace}

\newcommand{\review}[1]{{#1}}

\title{\systemname: A Dataflow-Based Framework for\\Human-AI Interactive Alignment in Urban Visual Analytics}

\author{%
  Gustavo Moreira, Leonardo Ferreira, Carolina Veiga, Maryam Hosseini, and Fabio Miranda
}

\authorfooter{
  \item Gustavo Moreira, Leonardo Ferreira, and Fabio Miranda are with the University of Illinois Chicago.
  	E-mail: \{gmorei3, lferr10, fabiom\}@uic.edu.
  \item Carolina Veiga is with the University of Illinois Urbana-Champaign. E-mail: carolvfs@illinois.edu.
  \item Maryam Hosseini is with the University of California, Berkeley. E-mail: maryamh@berkeley.edu.
}

\abstract{%
With the growing availability of urban data and the increasing complexity of societal challenges, visual analytics has become essential for deriving insights into pressing real-world problems.
However, analyzing such data is inherently complex and iterative, requiring expertise across multiple domains.
The need to manage diverse datasets, distill intricate workflows, and integrate various analytical methods presents a high barrier to entry, especially for researchers and urban experts who lack proficiency in data management, machine learning, and visualization.
Advancements in large language models offer a promising solution to lower the barriers to the construction of analytics systems by enabling users to specify intent rather than define precise computational operations. 
However, this shift from explicit operations to intent-based interaction introduces challenges in ensuring alignment throughout the design and development process. 
Without proper mechanisms, gaps can emerge between user intent, system behavior, and analytical outcomes.
To address these challenges, we propose Urbanite, a framework for human-AI collaboration in urban visual analytics. 
Urbanite leverages a dataflow-based model that allows users to specify intent at multiple scopes, enabling interactive alignment across the specification, process, and evaluation stages of urban analytics.
Based on findings from a survey to uncover challenges, Urbanite incorporates features to facilitate explainability, multi-resolution definition of tasks across dataflows, nodes, and parameters, while supporting the provenance of interactions.
We demonstrate Urbanite's effectiveness through usage scenarios created in collaboration with urban experts.
\systemname is available at \href{https://urbantk.org/urbanite}{urbantk.org/urbanite}.

}

\keywords{Urban analytics, urban data, dataflow, large language models, visualization framework, visualization system.}

\teaser{
  \centering
  \includegraphics[width=1\linewidth]{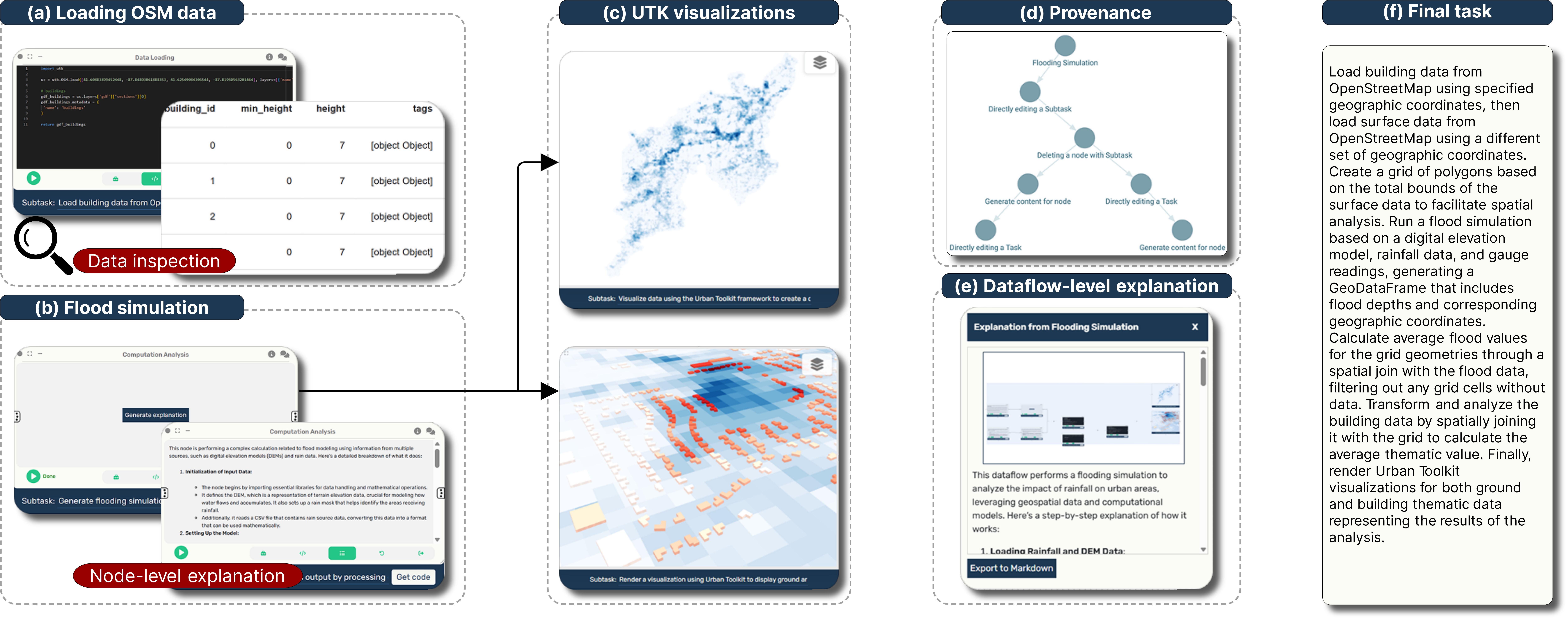}
  \caption{Using \systemname to analyze flood simulations. (a) The user loads a building layer and examines outputs through \textbf{provenance and data inspection}. (b) They create flood simulation nodes for a Chicago-area town and document them using \textbf{dataflow- and node-level explanations}. (c) Results and impacted buildings are visualized using UTK nodes. (d) The user modifies simulation parameters, branching the dataflow with \textbf{provenance and data inspection}. (e) To enhance documentation, they describe the dataflow using \textbf{dataflow- and node-level explanations}. (f) The final task is generated via \textbf{task \& subtask definition}, summarizing the dataflow.}
  \label{fig:teaser}
}

\graphicspath{{figs/}{figures/}{pictures/}{images/}{./}} 

\usepackage{tabu}                      
\usepackage{booktabs}                  
\usepackage{lipsum}                    
\usepackage{mwe}                       

\usepackage[hyphens]{url}
\usepackage{amssymb}
\usepackage{pifont}

\usepackage{mathptmx}                  
\usepackage{amsmath}
\usepackage[dvipsnames,svgnames]{xcolor}
\usepackage{soul}
\usepackage{tcolorbox}

\usepackage{titlesec}
\titlespacing*{\section}
{0pt} 
{*1.0} 
{*0.4} 

\titlespacing*{\subsection}
{0pt} 
{*0.8} 
{*0.6} 

\titlespacing*{\subsubsection}
{0pt} 
{*0.8} 
{*0.6} 

\newcommand{\hide}[1]{{}}
\renewcommand{\paragraph}[1]{\noindent\textbf{{#1}.}}

\newcommand{\circl}[2]{\tikz[baseline=-0.05in] \node[draw=none, fill=#1, text=white, circle, draw=black, ultra thin, minimum height=0.5em, inner sep=1pt,align=center,font=\sffamily\scriptsize,anchor=center,text height=1ex,text depth=.25ex] (Y) {#2};}

\newcommand{\rectangle}[2]{\tikz[baseline=(Y.base)] \node[draw=none, fill=#1, text=white, rounded corners=2pt, minimum height=1em, inner sep=1pt,align=center,font=\sffamily\scriptsize,anchor=center,text height=1.2ex,text depth=0ex] (Y) {#2};}

\definecolor{dataflowcolor}{HTML}{b8373e}
\definecolor{nodecolor}{HTML}{377eb8}
\definecolor{parametercolor}{HTML}{7eb837}
\definecolor{gray}{HTML}{737373}
\definecolor{purple}{HTML}{5D3FD3}

\newcommand{\dataflow}[1]{\rectangle{dataflowcolor}{#1}}
\newcommand{\node}[1]{\rectangle{nodecolor}{#1}}
\newcommand{\parameter}[1]{\rectangle{parametercolor}{#1}}
\newcommand{\expert}[1]{\rectangle{purple}{#1}}
\newcommand{\el}[1]{{\selectfont\sffamily\scriptsize{\textbf{#1}}}}

\begin{document}


\firstsection{Introduction}

\maketitle


The growing availability of urban data, coupled with increasing societal challenges, has driven the need for systems to support data-driven tasks across a range of domains, including urban planning, architecture, environmental and climate sciences, and public health.
Urban visual analytics (VA) systems have been repeatedly shown to be a key component in supporting these tasks~\cite{zheng2016visual,feng2022survey,deng2023survey,miranda2024state, ferreira2024assessing}, distilling complex data analytics workflows into accessible visual interfaces.
These systems enable urban experts to gain insights~\cite{lyu2023if}, detect patterns~\cite{garcia2019crimanalyzer}, or evaluate potential urban interventions~\cite{miranda2018shadow}.
%
%
%
Despite their growing importance, urban VA systems are still complex and resource-intensive, requiring expertise in urban science and multiple areas of computer science. Beyond technical challenges, their design is an iterative, collaborative process that must align expectations across diverse stakeholders~\cite{akbaba2023troubling}.

Recently, the visualization community has been highlighting tensions and technical hurdles in the process of authoring these systems~\cite{janicke2020visgap, akbaba2023troubling, wu2023grand, wu2023defence}.
These can be broadly categorized into three key interconnected areas.
First, urban data is inherently complex, encompassing heterogeneous, multi-scale, spatiotemporal datasets from diverse sources such as sensors, satellite and street-level imagery, and authoritative records.
The domain tasks are equally intricate, involving multiple coordinated steps~\cite{ferreira2024assessing, miranda2024state}. These complexities can stiffen the iterative and ideation cycles essential to visualization design~\cite{munzner2014visualization}, constraining flexibility and slowing down exploration of the visual analytics design space.
Second, developing VA systems demands diverse technical skills, often creating barriers for interdisciplinary teams. Domain experts may lack visualization expertise, while visualization experts may not fully understand domain-specific workflows, resulting in misaligned expectations and slower iteration. 
These knowledge gaps limit domain experts' contributions, diminishing the richness of collaboration by overlooking the valuable insights they can offer during the creative and conceptual phases of visualization design.
%
%
Third, developing such systems poses a high barrier to entry, and although urban experts are increasingly adopting data-driven approaches, they still face challenges in contributing to their design~\cite{wu2023grand, akbaba2023troubling}. 
Consequently, they are often treated as end-users rather than active participants in system development.
%
%
To address these challenges, many approaches now use low- or no-code paradigms to support VA~\cite{sicat2019dxr, gosling_lyi_2022, mcnutt2023no, moreira_urban_2024, l2024learnable}. 
However, these frameworks often lack the flexibility needed for complex urban workflows and the usability required for meaningful expert involvement.

Recent advancements in large language models (LLMs) provide an opportunity to further tackle these challenges by allowing users to focus on specifying intent rather than computational operations, helping non-technical users articulate goals without programming expertise.
However, this shift from explicit operations to intent-based interaction introduces challenges in ensuring alignment throughout the design and development process. 
Without proper mechanisms, gaps can emerge between user intent, system behavior, and analytical outcomes.
For example, LLMs can generate plausible but incorrect outputs~\cite{steyvers2025large}, and they lack transparency and structured mechanisms for iterative refinement of results.
Addressing these challenges requires new approaches that combine the strengths of LLMs with more structured design mechanisms, giving users the control to iteratively refine processes while benefiting from AI-assisted intent specification.
Additionally, the complex and iterative nature of these processes calls for tracking mechanisms to help users explore alternative design scenarios and recall their own process. Previous frameworks leveraged provenance~\cite{silva2007provenance, walker2013extensible, spinner2020explAIner, moreira2024curio} as an alternative to provide a structured and easy-to-access approach to explore user actions and, in this case, interactions with the LLM.



To address these challenges, we propose \systemname, a framework for human-AI collaboration in urban VA.
Conceptually, we ground the design of \systemname in the need for user-centered human-AI alignment across three objectives, as recently proposed by Terry et al.~\cite{terry2023interactive}: specification (what the AI should do), process (how the AI should do it), and evaluation (verifying and understanding what was done).
We also incorporate insights from a survey of experienced system builders, identifying key pain points in designing and developing urban VA systems.
Building on the theoretical framework and empirical findings, \systemname integrates novel features designed to enhance human-AI collaboration in urban VA.
At its core, \systemname follows a dataflow model, structuring systems as a series of nodes and edges that represent modular and interoperable VA components.
In other words, rather than treating VA systems as static software artifacts, we treat them as dataflows that can be iteratively constructed, adapted, and refined.
Such an approach enables the creation of lightweight, task-specific ``VA-lite'' systems.
\systemname's dataflows are organized around a specification language, which ensures consistency in defining and refining analytical workflows.
Additionally, this language serves as a layer between the LLM and the user, constraining the model's output to ensure alignment with analytical goals and system constraints.
To support alignment between user intent and system behavior, \systemname introduces design features that operationalize the alignment principles.
First, it enables users to specify their analytical goal in natural language through a series of guided prompts that result in a task, subtasks, and a dataflow sketch.
Second, it generates contextualized code to fulfill the predefined task and implement the sketch, as well as continuous node and edge suggestions to guide the process. 
Third, explainability and provenance mechanisms support the evaluation of the final result and the exploration of alternative scenarios.
\review{Our contribution also lies in demonstrating how LLMs, embedded within a dataflow system, can enhance urban VA by bridging the semantic gap between the complex (and often ill-defined) intents of experts and the formal, structured requirements of VA systems in this domain. Specifically, we show how such integration lowers the barrier for domain experts to translate high-level goals into executable workflows, facilitates iterative exploration while maintaining transparency, and enables human-AI alignment through the dataflow acting as an interpretable intermediary between LLM outputs and expert understanding.}
\review{We evaluate \systemname through usage scenarios, experts' feedback, and a quantitative evaluation.}
\systemname is available at \href{https://urbantk.org/urbanite}{urbantk.org/urbanite}.
\section{Related Work}

%
%

\subsection{Urban visual analytics systems}

Urban VA combines computational methods and interactive visualization to help experts analyze complex urban data~\cite{zheng_urban_2014}.
%
%
Recent surveys highlight the field's growth and complexity~\cite{zheng2016visual,adrienko2017visual,feng2022survey,deng2023survey, miranda2024state, ferreira2024assessing}.
Recently, Deng et al.~\cite{deng2023survey} surveyed over 200 visualization papers, focusing on how different models and analysis approaches are integrated in the visualization workflow, while our recent reviews surfaced over 450 papers on 3D urban data~\cite{miranda2024state} and over 130 papers with requirements for urban-specific frameworks~\cite{ferreira2024assessing}.
%
%
%
%
These works have demonstrated the importance of supporting diverse stakeholders, including urban planners, architects, social scientists, climate researchers, public health professionals, and community advocates, each with distinct analytical needs and priorities.
Despite the proliferation of urban VA tools, they still face fundamental challenges, particularly their difficulty in construction and their siloed nature, as detailed in Section~\ref{sec:background}.
As a result, urban VA remains fragmented, with many systems operating in isolation rather than as part of a unified ecosystem.

Given these challenges, experts increasingly turn to computational notebooks, which provide an accessible, flexible, and reproducible way to analyze urban data without the burden of developing full-fledged VA systems.
However, while notebooks offer ease of construction and experimentation~\cite{lau2020design}, they lack the domain-specific capabilities, scalability, and interactivity of bespoke systems.
Addressing these challenges requires a shift towards more modular, extensible, and adaptable frameworks that can bridge the gap between bespoke applications and broadly applicable urban analytics solutions.
In our previous work, Curio~\cite{moreira2024curio}, we focused on enabling human-human collaboration by supporting teams of experts in jointly exploring and analyzing data.
With \systemname, we extend this vision to human-AI collaboration, leveraging a natural language interface that allows users to engage with the system more intuitively.
%
\systemname combines the flexibility of custom VA systems with accessibility, modularity, and reproducibility, letting experts build and share analyses without traditional development overhead.
By integrating an AI assistant through LLMs, \systemname further lowers barriers by assisting users in navigating complex workflows, automating data steps, and generating visualizations.

\subsection{LLM- and NLI-enhanced visual analytics}


%

LLMs have recently proven effective for tasks like text summarization, report generation~\cite{minaee2025largelanguagemodelssurvey}, and programming support~\cite{chen2021evaluating}.
Their integration with natural language interfaces in VA is emerging as a way to reduce barriers to complex data analysis.
Shen et al.~\cite{shen2023towards} provided a comprehensive survey of NLI applications in visualization, outlining emerging trends and challenges, including provenance tracking of prompts for interpretability.
Basole and Major~\cite{basole2024generative} further explored the role of generative AI and natural language prompts across the visualization workflow, identifying key opportunities and obstacles in integrating AI-driven assistance into analytical processes.
%
%
While prior studies highlight broad challenges in integrating NLIs and LLMs into visualization workflows, Terry et al.~\cite{terry2023interactive} examine the alignment of human and AI capabilities within interactive systems. Their work proposes a conceptual framework for effective human–AI integration, complementing more focused efforts that embed LLMs directly into the visualization process.
Zhao et al.~\cite{zhao2024leva} proposed using LLMs to support VA across multiple stages of user workflows, including onboarding, exploration, and summarization. They also introduced a specification framework for VA systems that enables LLMs to interpret visual views and their interrelationships.
Tian et al.~\cite{tian2024chartgpt} explored the use of LLMs for generating charts, proposing a step-by-step reasoning pipeline that decomposes visualization tasks into more manageable sub-tasks.
Tang et al.~\cite{tang2024steering} investigated the use of LLMs for summarization using visual workspaces as a steering mechanism.
%
%
Sah et al.\cite{sah2024generating} used LLMs to generate visualization specifications by detecting data attributes, inferring tasks, and recommending visualizations.
%
L'Yi et al.\cite{l2024learnable} built a visualization system that combines multimodal interfaces, including natural language, to improve usability.

We draw upon research on alignment in interactive systems~\cite{amershi2019guidelines, yang2020re, terry2023interactive}, particularly the shift towards the specification of outcomes rather than operations.
\systemname builds on these principles by introducing new mechanisms for aligning AI assistance with user intent in urban analytics.
Influenced by Zhao et al.~\cite{zhao2024leva}, we also introduce an urban VA-specific specification for the creation of systems, which acts as a mediator between the AI and the user prompt.
\systemname embeds LLM guidance \review{into} structured dataflows, enabling transition between natural language interaction and interactive visual exploration.

\subsection{Dataflow-based approaches in visual analytics}

The use of dataflows as a structured paradigm for data analysis has long been explored across various topics, including database~\cite{silva2018dfanalyzer,sun2023bigdataflow}, information visualization~\cite{elmqvist2007datameadow, javed2013explates, yu2016visflow}, scientific visualization~\cite{callahan2006vistrails, waser2011nodes}, and visualization education~\cite{silva2011using}.
%
%
%
Systems employing this paradigm use diagrams with nodes \review{representing} data transformations and visualizations, and edges \review{defining the} dataflow. Users then interactively compose data pipelines to specify their intent, often through drag-and-drop interfaces that allow visual construction and modification without requiring extensive programming expertise.
Importantly, dataflow systems promote transparency, modularity, and reusability, allowing analytical components to be adapted across different usage scenarios~\cite{moreira2024curio}.
Recent works have extended the applicability of dataflows within the context of VA.
Ulbrich et al. introduced sMolBoxes~\cite{ulbrich2022smolboxes}, a dataflow model designed for molecular dynamics exploration, demonstrating how structured dataflows can support specialized scientific tasks.
Yu and Silva presented VisFlow~\cite{yu2016visflow}, a framework that supports the creation of interactive charts within a dataflow. The framework was later extended into FlowSense~\cite{yu2020flowsense} by incorporating a natural language interface featuring a semantic parser that recognizes specific utterances within the dataflow environment. 
%
Curio~\cite{moreira2024curio} builds upon these principles by introducing a collaborative dataflow environment. Unlike traditional dataflow systems that primarily focus on individual data pipeline creation, Curio facilitates collaboration through discussion spaces, annotation mechanisms, and provenance tracking.

Building on Curio, our system enhances dataflow-based VA by integrating AI-assisted guidance for urban analytics.
While Curio focused on human-human collaboration, our system introduces human-AI collaboration, allowing users to describe analytical intent in natural language, which is then translated into structured dataflows.
%

\section{Background: Challenges in Urban Visual Analytics}
\label{sec:background}


Urban analytics has emerged as a valuable approach to support decision-making in urban contexts, enabling domain experts to explore and analyze diverse urban data sources~\cite{miranda2016urban, li2019cope, deng2019airvis, garcia2019crimanalyzer, miranda2020urban, dong2024tcevis}.
These experts -- such as urban planners, engineers, climate scientists, and public health professionals --  primarily rely on three environments to support these analyses: off-the-shelf tools, computational notebooks, and urban VA systems.
Off-the-shelf tools, such as ArcGIS, offer standardized functionalities but often come with steep learning curves~\cite{ziegler2023need}.
Computational notebooks require programming skills~\cite{wang2021integrating, yap_free_2022}.
In contrast, urban VA systems simplify workflows into interactive tools to analyze data. 
However, development remains challenging due to the complexity of heterogeneous, multimodal data, and the need to support collaboration among stakeholders with diverse expertise and priorities.
%
%

These challenges complicate the creation of broadly applicable VA systems suited for real-world urban settings.
For example, large-scale streaming and image data often require intensive wrangling or costly feature extraction, slowing iteration and locking in early design choices\review{.}
%
%
%
Without standardized, reusable pipelines, efforts are often duplicated, leaving most urban VA systems bespoke and narrowly tailored.
%
The visualization community has recognized this foundational challenge~\cite{wu2023grand}, but it is especially pronounced in urban VA due to unique contextual constraints. 
%
%
Unlike other domains with standardized data formats, urban data varies by region due to local practices and policies. 
Combined with the black-box nature of bespoke systems, this restricts their adaptability and keeps the broader applicability of urban VA systems limited.
%
%
%
%
As a result, urban experts struggle to build on prior work, and insights from one context rarely transfer, reinforcing fragmentation.

Recognizing these limitations, recent years have seen a push towards more modular and adaptable urban VA frameworks~\cite{salinas2022cityhub, moreira2024curio, chen2024sensemap, moreira_urban_2024}.
Inspired by toolkits and frameworks in other domains~\cite{buschel2021miria, tiankai2022fairrankvis, qiu2024docflow}, researchers have begun decomposing complex systems into modular building blocks, offering greater reusability and extensibility. \review{This modular decomposition works well because VA systems can be modeled and interpreted, fundamentally,  as complex dataflows.}
With Curio~\cite{moreira2024curio}, we took a first step towards this vision with a dataflow-based framework grounded in a knowledge base of reusable VA components, primarily supporting human-human collaboration between visualization researchers and urban experts.
Curio bridges monolithic VA systems and computational notebooks by making operations transparent to both urban experts and visualization researchers. However, its dataflow and drag-and-drop interface still demand that users understand data operations and manually build workflows.
%
%
%
LLMs present a promising avenue to overcome these barriers. By enabling natural language interactions for the authoring of VA components, LLMs provide domain experts with a more intuitive way to specify their intent, facilitating expert-AI collaboration.
Unlike traditional interfaces that bridge the gaps of execution and evaluation, however, LLM-based systems introduce a qualitatively different interaction paradigm: \emph{intent-based outcome specification}~\cite{nielsen2023ai}.
Instead of specifying operations, users describe desired outcomes, requiring well-designed human-AI interfaces to bridge additional alignment gaps in specification, process, and evaluation~\cite{terry2023interactive}.
%
%
Our work designs mechanisms to bridge human-AI collaboration gaps, leveraging LLMs for multi-level intent specification. This enables urban experts to define goals, implement dataflows to fulfill them, and understand the process via automated documentation and provenance.

\section{Understanding Urban Visual Analytics Challenges}
\label{sec:survey}

%
With \systemname, we aim to simplify dataflow creation for urban VA by bridging technical complexity and domain expertise.
As a step towards this vision, we were particularly interested in exploring what it would mean to have an AI-enhanced system that assists urban experts in authoring their own VA interfaces.
%
%
%
While our objective centers on supporting urban experts directly, our goal with this survey was to understand, from the perspective of visualization experts, the key gaps and collaborative dynamics that shape the development of urban VA systems. 
%
%
%
%
We explored how teams bridge expertise, define goals, document work, and coordinate across design phases. By gathering insights from recent projects, we aimed to surface challenges, best practices, and opportunities to guide \review{Urbanite’s} design.
%
%
%

\myparagraph{Survey design and participants}
%
The survey had two parts.
%
%
First, participants reflected on working with urban experts, covering project duration, code availability, collaboration barriers, iteration frequency, expectations, and documentation practices.
%
%
Then, participants estimated time spent across four design stages (Understand, Ideate, Make, Deploy), following  McKenna et al.~\cite{mckenna2014design}, describing key activities, tools, and decisions.
%
%
%
We recruited 10 participants, targeting first or second authors of urban VA system papers, based on our prior surveys~\cite{ferreira2024assessing, miranda2024state}. The group included 2 faculty members, 4 PhD students, 3 research staff, and 1 postdoc.
\review{We focused on visualization experts, as they are part of the VA system-building processes we want to support.}


Next, we synthesize the key themes identified through our survey.

\myparagraph{Time spent in each design stage}
%
%
On average, participants spent 24\% on Understand, 23\% on Ideate, 33\% on Make, and 20\% on Deploy. Notably, two spent ~40\% on Understand (deep exploration), while two others spent ~40\% on Deploy (system refinement).

\myparagraph{Bridging the expertise gap between domains}
Five participants reported misunderstandings or communication breakdowns between urban experts \review{(e.g., architects, urban planners)} and visualization experts.
%
%
The most cited barrier (40\%) was unclear or incompatible terminology between visualization and domain experts. Another 30\% reported multiple barriers, including technical limitations, expertise gaps, differing methods, and communication issues.

%

\myparagraph{Narrowing down analytical goals}
Some participants (40\%) experienced evolving or confusing goals, highlighting variability in how analytical objectives were established.
Reflecting on their iterative approach, one participant noted: ``I began by implementing simple, easy-to-understand visualizations based on my understanding of the data. I then consulted domain experts to gather insights and feedback. Using their feedback, I experimented with combining multiple visualizations into single views.''
While most participants felt the iterative process of refining their systems was as expected (60\%), 40\% experienced either more or fewer iterations than anticipated, pointing to inconsistent expectations about the collaborative design process.

\myparagraph{Tool availability and dissemination}
Survey responses revealed that open-source dissemination of urban VA systems remains limited. Only one participant explicitly indicated that both the system and its code were made publicly available in a way that qualifies as open source. Two others shared only the system interface. Four participants reported that neither the system nor the code was made publicly accessible, typically due to institutional, privacy, or operational constraints. Fully open dissemination of tools is still relatively uncommon, limiting opportunities for reuse, validation, and community uptake.

\myparagraph{Provenance of analyses \& artifacts}
Most participants (60\%) had clearly defined project goals, while 40\% faced evolving or unclear ones -- highlighting the difficulty of setting consistent objectives in interdisciplinary work. 
%
%
Documentation practices varied considerably, with manual documentation most commonly used. 
%
%
One respondent noted that much of the work involved ``unglamorous software engineering.''
Dedicated provenance tools, such as version control or specialized tracking systems, were utilized infrequently. 
One participant described a more structured approach that ``at various stages (...) stakeholders were shown the current version of the system.''

\myparagraph{Key takeaways}
The survey highlights that the most significant challenge across projects was bridging domain expertise, primarily due to persistent communication and terminology issues between urban and visualization experts.
Participants mentioned the key role of defining analytical goals, noting that projects with ambiguous or evolving objectives often resulted in inefficiencies and misaligned expectations.
Furthermore, the results indicate gaps in documentation practices, with many teams relying on informal or manual methods rather than structured provenance tracking.
Overall, addressing communication barriers, refining goal-setting practices, and improving documentation emerged as central themes for enhancing outcomes in urban VA.



\section{The \systemname Framework}

Building on the challenges and design opportunities identified in previous work and our survey, we present \systemname, a framework designed to support the authoring of modular, transparent, and adaptable VA workflows for urban data.
\systemname addresses the limitations of current approaches, including the opacity of monolithic systems and the technical barriers of computational notebooks, by enabling users to co-construct dataflows using reusable components and intuitive interfaces.
A central aim of \systemname is to lower the barriers that often prevent urban experts from directly engaging in the creation and adaptation of VA systems.
To this end, \systemname integrates an LLM to enable intent-based interactions throughout the authoring process, which is modeled as a dataflow.
\review{This dataflow serves as a transparent intermediary, allowing users to understand and validate the LLM's interpretation. It explicitly supports a human-in-the-loop paradigm, supporting users to iteratively refine LLM suggestions and data transformations. Furthermore, dataflow components are inherently modular and reusable, enhancing efficiency and consistency across projects.}
%
To facilitate meaningful human-AI alignment, we introduce a framework for multi-level intent specification, operationalizing this alignment across scopes and modes of expression. 
By allowing users to specify intent and interpret system responses at varying levels of specificity, \systemname supports AI-driven assistance that is both traceable and adaptable to the user’s analytical context.
\review{\systemname leverages conceptual and technical foundations laid by UTK~\cite{moreira_urban_2024} and Curio~\cite{moreira2024curio} while incorporating an LLM-based intent translation pipeline. See supplementary material for key differences between \systemname and these frameworks.
}
We first introduce the framework’s theoretical foundations (Section~\ref{sec:foundations}) and design goals (Section~\ref{sec:goals}). We then present the interaction space for building dataflows  (Section~\ref{sec:interactions}), detail the multi-level authoring mechanisms (Section~\ref{sec:features}), followed by an overview of the system (Section~\ref{sec:system}).



\subsection{Conceptual foundations}
\label{sec:foundations}

\systemname is built on the premise that human-AI alignment can be achieved through thoughtful interaction design.
Specifically, it draws from the conceptual framework of \emph{interactive alignment} recently introduced by Terry et al.~\cite{terry2023interactive}, which reconceptualizes the classic human-computer interaction cycle in light of modern, intent-based AI systems.
%
%
Traditional interfaces rely on direct manipulation, requiring users to choose and sequence actions to reach their goals.
%
In contrast, human-AI interaction uses a declarative approach: users state desired outcomes, and the system interprets and executes them. This lowers entry barriers but introduces ambiguity around system capabilities, behavior, and intent alignment.
\review{Terry et al.~\cite{terry2023interactive} propose} three distinct stages to support more transparent and effective interaction in a declarative paradigm.
We adopt their conceptual framework to operationalize the human-AI interaction cycle in the context of dataflow authoring through three key alignment objectives:

\myparagraph{(1) Specification alignment} \emph{What the dataflow should accomplish} -- that is, users' analytical intent and goals (e.g., ``identify areas with high heat vulnerability''), and how these translate into system behavior; 

\myparagraph{(2) Process alignment} \emph{How the dataflow should be constructed or operations performed} -- the data transformations, analytical operations, and visualizations that the AI proposes to fulfill users' intent; 

\myparagraph{(3) Evaluation alignment} \emph{Verifying \& refining the dataflow and outputs} -- ensuring that the responses are correct and meet users' needs.

In \systemname, this alignment framework directly informs our system architecture and design.
A user's analytical goals are translated into data pipelines composed of data transformations, analytical operations, and visualizations.
Interaction mechanisms at multiple levels of granularity and abstraction then guide users through the authoring, refining, and evaluation of dataflows, ensuring that alignment is always met.

\subsection{Design goals}
\label{sec:goals}

Building on the previously identified challenges in urban VA and the conceptual foundations of human-AI alignment, we articulate a set of design goals that guide the development of \systemname.
These design goals (DGs) are directly informed by the real-world challenges described in Section~\ref{sec:background} and the empirical findings from our survey in Section~\ref{sec:survey}.
In particular, the high technical barriers to authoring systems (DG1, DG2, DG6), the need for transparency (DG3, DG4), and the evolving nature of urban VA workflows (DG5, DG6).
%
Each design goal reflects a specific response to misalignment risks in \textbf{specification} (S), \textbf{process} (P), \textbf{evaluation} (E), or all of them (cross cuts, CC).

\myparagraph{\rectangle{black}{DG1-S} Enable natural language specifications of modular dataflows}
Dataflows are inherently modular, composed of discrete components for data processing, analytical operations, and visualization.
However, translating analytical goals into these components is a key barrier for domain experts.
\systemname must allow users to specify intent in natural language.
Through LLM integration, \systemname must interpret these goals and map them to partial or complete dataflows, reducing reliance on technical expertise.

\myparagraph{\rectangle{black}{DG2-P} Support user control during dataflow authoring}
Users must be able to control \emph{how} the framework translates goals into operations.
Rather than treating the dataflow generation as a \review{black box}, \systemname must provide access to intermediate decisions made by the LLM, and users should accept, reject, or revise these decisions.
Control must be supported through natural language (e.g., ``filter the data and perform a spatial join'') and manipulation (e.g., dragging a module into place).

\myparagraph{\rectangle{black}{DG3-E} Enable inspection of AI-generated outputs}
Outputs generated by LLMs and other AI components must be made transparent and inspectable to support evaluation alignment. \systemname must allow users to verify whether the system’s responses align with their original intent, not only in terms of final results but also in how those results were derived.
For example, if a user asks ``visualize the most heat-vulnerable neighborhoods using census and weather data'', \systemname must be able to not only inspect a heatmap, but also which datasets were used and how the vulnerability index was computed.

\myparagraph{\rectangle{black}{DG4-E} Trace and visualize the provenance of data transformations and human-AI interactions}
\systemname must expose the provenance of data transformations and LLM outputs throughout the lifecycle of the dataflow.
This must support the evaluation and comparison of different states of the dataflow, as well as the user prompts used to generate them.
%
%
For example, if an expert is exploring accessibility datasets, they may ask the system ``highlight neighborhoods that are outliers in terms of accessibility problems.'' Over time, the expert might modify the definition of outlier or adjust the spatial resolution of the analysis. Provenance tracking must allow the expert to see which revision introduced changes, reuse, or revert earlier steps without reconstructing the dataflow from scratch.


\myparagraph{\rectangle{black}{DG5-CC} Support iterative refinement across alignment stages}
Authoring is inherently iterative: goals evolve, understanding deepens, and constraints shift over time.
To maintain alignment between user intent and AI behavior, \systemname must support refinement across all stages -- specification, process, and evaluation.
This includes the ability to revisit and revise prior steps: rephrasing analytical goals, modifying AI-generated dataflow components, or adjusting outputs based on observed results.
These iterations should be fluid, allowing users to go back and forth between stages as needed.
\systemname must support these refinements through both natural language and direct manipulation.

\myparagraph{\rectangle{black}{DG6-CC} Support multi-level interaction within alignment stages}
While DG5 focuses on iteration across stages, DG6 focuses on depth within each stage.
Given the range of user expertise and task complexity, \systemname must support interaction at multiple levels of abstraction within specification, process, and evaluation.
For example, during specification, users may articulate a high-level objective (``analyze heat vulnerability'') or fine-tune specific prompts; during process, they might inspect the dataflow or dive into the inner logic of individual modules; during evaluation, they may explore summaries or individual outputs.

\subsection{A flexible interaction space for building dataflows}
\label{sec:interactions}

In \systemname, the user's primary goal is to construct a dataflow: a data pipeline of analytical operations that serves as a model of an urban VA system.
The authoring revolves around two core responsibilities: (1) defining the structure of the dataflow (i.e., breadth), and (2) specifying or modifying the inner workings of each module within the structure (i.e., depth).
Users may begin by sketching out a pipeline, either through natural language or by manually composing dataflow nodes. This defines the high-level flow of data and operations.

However, authoring is rarely linear. As users dive into the details of a specific module, they may realize that the structure itself needs to change.
Likewise, while building out the structure, users may iteratively define or revise the behavior of individual nodes to test their ideas or explore alternative paths.
To support this interaction space in \systemname, we define a two-dimensional conceptual framework with the \emph{scopes of specification} and \emph{modes of expression}.
The scopes of specification refer to the granularity of \emph{what} is being authored -- ranging from the entire dataflow to individual parameters within a module. The modes of expression capture \emph{how} that specification is articulated -- whether through natural language, visual manipulation, declarative grammar, or executable code.
%
%
%
Changes made at one scope with one mode are reflected across others. For instance, updating a parameter in a node's code will be reflected in the natural language explanation.
\systemname continuously synchronizes these representations, giving users the flexibility to shift between modes and scopes while preserving consistency and intent.
%
%
Table~\ref{tab:scope-mode-expression} and Figure~\ref{fig:features} summarize the supported interactions and how they contribute to \systemname's integrated, multi-modal authoring experience.
The specific design features that support these dimensions in practice are presented in Section~\ref{sec:features}.



\subsubsection{Dataflow preliminaries}

\systemname's dataflow model follows from the one first introduced in our Curio framework~\cite{moreira2024curio}.
A dataflow is defined as a composition of modular computing nodes, connected through data and interaction dependencies, and designed to express VA workflows. 
%
%
%
Each data node takes as input and produces zero or more data layers.
Although the formalism treats all nodes uniformly, each node can semantically represent a distinct operation type within the workflow.
\systemname adopts the types defined in Curio, including data wrangling and transformation, analysis and modeling, and visualization nodes. 
To reduce ambiguity in how goals are interpreted and operationalized, \systemname introduces a structured \emph{specification layer} that builds directly on this formal model.
This specification serves as a machine-readable representation of the current dataflow and provides a bridge between high-level user intent and low-level implementation. 
It enables the system to capture both the evolving scope of specification -- from entire workflows to individual parameters -- and the mode of expression -- whether defined through natural language, UI, declarative grammar, or executable code.

%
While Curio used this dataflow model for collaborative execution, \systemname repurposes it as the backbone for LLM-assisted, multimodal authoring.
It serves both as an execution plan and as the foundation for prompt-based refinement, traceable suggestions, and semantic alignment. \review{The dataflow also acts as a transparent intermediary between user intent and LLM output, supporting in-the-loop refinement and reusable components.}
%
%
\review{See supplementary material for full specification details.}
%
%
%



\begin{table}[t!]
\centering
\caption{Combinations of specification scopes and expression modes.}
\vspace{-.25cm}
\begin{tabular}{p{3.8cm}|cccc}
\multicolumn{1}{c}{} & \multicolumn{4}{c}{\textbf{Modes of expression (\emph{how})}} \\
\cmidrule(lr){2-5}
\textbf{Scopes of specification (\emph{what})} & \textbf{NL} & \textbf{UI} & \textbf{Grammar} & \textbf{Code} \\
\midrule
Dataflow  & \circl{dataflowcolor}{ } & \circl{dataflowcolor}{ } & \circl{dataflowcolor}{ } &  \\
Module    & \circl{nodecolor}{ } & \circl{nodecolor}{ } & \circl{nodecolor}{ } & \circl{nodecolor}{ } \\
Parameter &  & \circl{parametercolor}{ } & \circl{parametercolor}{ } & \circl{parametercolor}{ } \\
\end{tabular}
\vspace{-.5cm}
\label{tab:scope-mode-expression}
\end{table}

\subsubsection{Scopes of specification}


We define the \emph{scopes of specification} as the granularity in which the user interacts with \systemname to create a dataflow.
This ranges from dataflow definitions that capture the overall structure of the data pipeline, to mid-level specifications of individual modules, down to fine-grained parameter settings within those modules.

\myparagraph{\circl{dataflowcolor}{ } \dataflow{Dataflow-level specification} }
At the dataflow level, users specify the overall structure of the workflow.
This includes defining the key \el{nodes} that will be part of the dataflow, the data \el{inputs} and \el{outputs}, and how these nodes are connected through data and interaction dependencies.
The result is a blueprint that outlines the logical flow of data and operations.
Operations at this level restrict interactions to coarse-grained structural elements: users can create, modify, or remove \el{nodes} and \el{edges}, as well as define or update the \el{task} that describes the dataflow's high-level task.
At this level, the dataflow structure provides context for downstream refinement and serves as a scaffold for LLM-driven suggestions and user guidance.

\myparagraph{\circl{nodecolor}{ } \node{Module-level specification}}
At the module level, users focus on specifying the behavior and configuration of a single node within the dataflow.
Each node is a discrete data, analytical, or visual operation.
At this level, users are restricted to modifying the internal contents of a node.
This includes the \el{subtask} that describes the node’s purpose, the \el{content} field that encodes the operation logic (e.g., grammar or code).
The structure of the dataflow remains unchanged at this level: users are not adding new modules or rerouting connections, but refining how a specific module contributes to the overall dataflow.

\myparagraph{\circl{parametercolor}{ } \parameter{Parameter-level specification}}
At the parameter level, users engage with fine-grained aspects of the dataflow behavior: individual settings and thresholds that define how components operate.
The parameters are made available to the user through UI elements using code annotations~\cite{moreira2024curio}.
When operating at this level, user modifications are restricted to the code annotations within a single node. The structure of the dataflow and the purpose of the node remain fixed.

\myparagraph{Provenance and interoperability across scopes}
All interactions, regardless of the scope, are captured and stored in a hierarchical data structure.
This interoperability is made possible by considering the dataflow specification as a central artifact.

\subsubsection{Modes of expression}

Scopes define \emph{what} part of the dataflow the user edits; \emph{modes of expression} define \emph{how} they convey intent. In \systemname, users can use natural language, UI, code, or grammar, with changes reflected across all modes through the shared specification.

\myparagraph{Natural language}
This is the most accessible entry point for authoring in \systemname.
Users can describe high-level tasks (e.g., ``map accessibility gaps across neighborhoods'') or specific prompts (e.g., ``map median income to color''). These are interpreted by the LLM and translated into modifications to the dataflow specification.
Natural language can be used across scopes, from defining dataflows to adjusting parameters.

\myparagraph{User interface}
%
The interface lets users build and edit dataflows by directly dragging nodes and adjusting parameters. Based on Curio’s components, it supports extensibility: users can add elements via code annotations as reusable widgets. The UI and diagram remain the specification’s anchor, whether authored visually or in code.

\myparagraph{Code \& grammar}
\systemname supports authoring via Python and declarative grammars (Vega-Lite, UTK~\cite{moreira_urban_2024}) to define node logic.
\review{Each node executes its logic using imported libraries, meaning \systemname's performance is tied to the efficiency of the specific libraries. This enables flexible and transparent node authoring, but can constrain performance for large datasets or complex visual encodings.}

\subsection{Design features for human-AI alignment}
\label{sec:features}

%
Building on the earlier foundation, \systemname implements features that align user intent with AI behavior during dataflow authoring.
These features support alignment across specification, process, and evaluation, contributing to key design goals and tied to the scopes of specification.
%
%
%

\subsubsection{Specification features}

\myparagraph{\circl{dataflowcolor}{ } Dataflow generation and scaffolding}
The process of dataflow creation in \systemname begins with a conversational interaction between the user and the LLM. Through this dialogue, the user articulates their analytical goals in natural language, \rectangle{black}{DG1-S}.
The system incrementally builds a structured task description from this input.
%
This occurs through guided prompts, with the LLM reformulating user input into a machine-readable schema and requesting clarification as needed.
This exchange supports \rectangle{black}{DG5-CC}.
To ensure the task is well-defined, \systemname performs schema checks that verify key elements (e.g., presence of a dataset, whether the task can be mapped to one or more supported nodes).
Once requirements are met, the system transitions from intent capture to dataflow construction.
At this point, \systemname uses the finalized specification to scaffold an initial dataflow.
%
%
This involves creating nodes for key analytical operations and edges for data or interaction flow. 
%
%
%
Each node is assigned a subtask (based on the main task) describing its role in natural language. Stored in metadata, it guides LLM assistance for code, grammar, and UI generation.
%
%

\begin{figure}[t!]
    \centering
    \includegraphics[width=\linewidth]{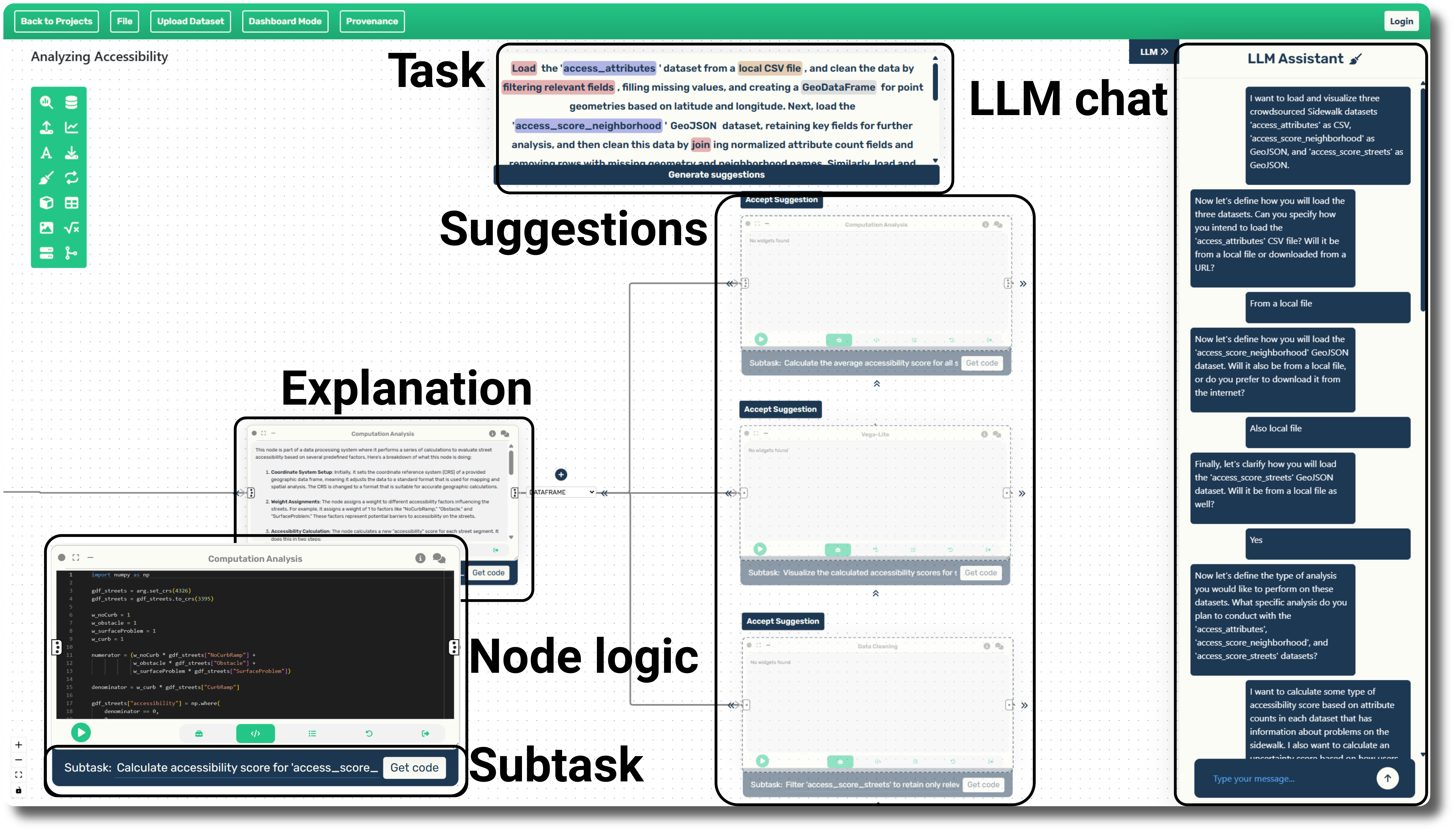}
    \caption{\systemname's interface, with design features for human-AI alignment.}
    \vspace{-.1cm}
    \label{fig:features}
    \vspace{-.5cm}
\end{figure}

\myparagraph{\circl{dataflowcolor}{ } \circl{nodecolor}{ } Task \& subtask definition}
After the initial dataflow is scaffolded through the conversational interaction with the LLM, \systemname assigns a natural language task description to the entire dataflow (supporting \rectangle{black}{DG1-S}), along with subtask descriptions to each node.
The task captures the user’s overarching analytical goal (e.g., ``analyze accessibility issues across neighborhoods''), while each subtask expresses the local intent of a node within that broader workflow (e.g., ``join accessibility data with demographic layers'').
Such a multi-level interaction within specification supports \rectangle{black}{DG6-CC}.
These descriptions are editable: users can revise the task to reflect updated goals or change the subtask to clarify, refine its behavior.
To ensure coherence across the dataflow, \systemname uses a two-way synchronization mechanism between task and subtasks.
When a user edits the task, \systemname triggers a decomposition routine that prompts the LLM to suggest revised subtasks for each node in the dataflow, further supporting \rectangle{black}{DG6-CC}. 
%
%
Conversely, when a user modifies a subtask, \systemname re-evaluates the full set of subtasks to determine whether the overall task description is still aligned.
%
%
This uses an LLM summarization pass to generate a natural language summary of current subtasks and check for coherence, implementation issues, or opportunities to refine or split subtasks. If found, a warning is added.
All synchronization steps are tracked via the provenance model.
%




\begin{figure*}[t!]
    \centering
    \includegraphics[width=1\linewidth]{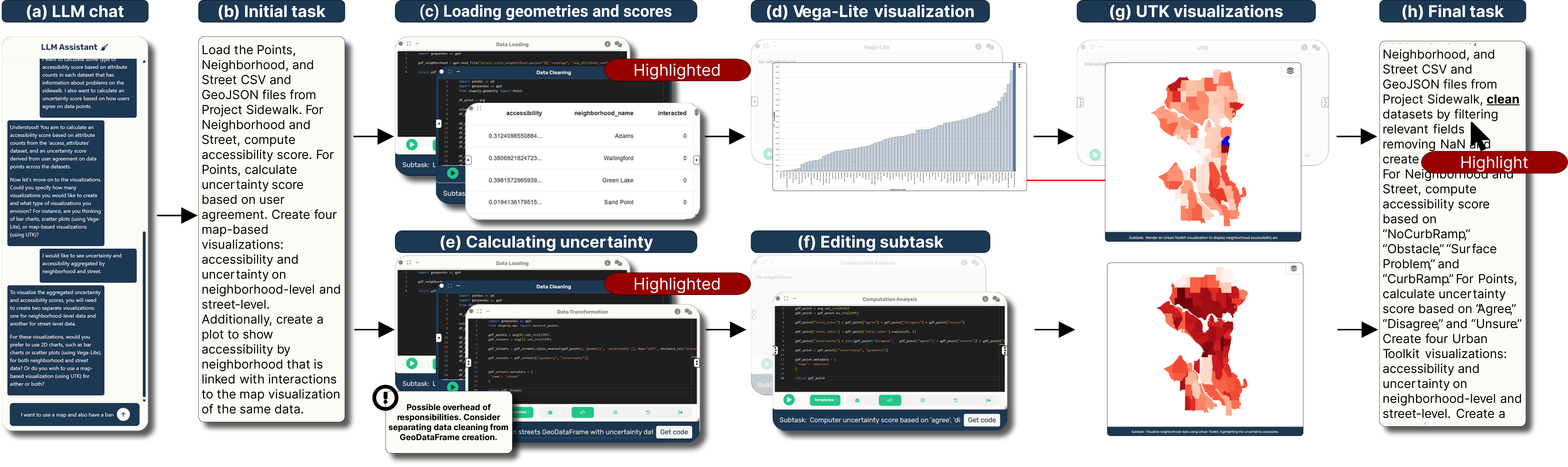}
    \vspace{-0.6cm}
    \caption{Using \systemname to analyze Project Sidewalk accessibility data. (a) The user starts with \textbf{dataflow generation} to define a task. (b) Based on the task, the LLM proposes a first sketch. (c) The user accepts nodes for geometry and score calculation, using \textbf{code generation} to fill them in. (d) The same is done for a neighborhood-level bar chart. (e) \review{In a parallel flow,} uncertainty is then calculated, and (f)  the user edits the subtask to aggregate by neighborhood using \textbf{task \& subtask definition}. (g) UTK suggestions are accepted to visualize accessibility (top) and uncertainty (bottom). (h) As nodes and subtasks were edited, the task was automatically refined. Hovering task keywords highlights related dataflow parts.}
    \label{fig:case1}
    \vspace{-0.6cm}
\end{figure*}

\subsubsection{Process features}

\myparagraph{\circl{nodecolor}{ } Code generation}
For each node, \systemname uses the natural language subtask as the primary prompt for generating the node's logic.
Depending on the node type, this logic may be rendered as Python code or declarative grammar (Vega-Lite or UTK).
The prompt is specified with both subtask text and contextual metadata (expected inputs and outputs, node type) to generate a candidate implementation.
Users remain in control throughout the process \rectangle{black}{DG2-P}. They may inspect, modify, or fully rewrite the generated logic.
Alternatively, the user can request \systemname to revise the implementation by updating the subtask.
%
To maintain consistency across abstraction levels, \systemname synchronizes edits made with different modes of expression, supporting \rectangle{black}{DG5-CC}.
When a user makes changes to the code or grammar, \systemname prompts the LLM to infer and update the corresponding subtask, avoiding drifting between the descriptive and functional representations of each node.
%


\myparagraph{\circl{dataflowcolor}{ } Connection suggestions}
%
%
As the user iteratively \review{builds} out their dataflows, they may reach decision points where the next analytical step is unclear or open-ended.
To support exploration in these moments, \systemname allows users to request connection suggestions from the LLM.
\systemname generates these suggestions by prompting the LLM with the current state of the dataflow specification, including existing nodes, their subtasks, and their interconnections.
\systemname then returns candidate nodes (e.g., new analysis, transformation, or visualization steps), each annotated with a subtask that \review{explains} its intended logic.
These proposals are presented as options rather than decisions, supporting \rectangle{black}{DG2-P}.
Users maintain full control over which nodes to integrate.


\subsubsection{Evaluation features}

\myparagraph{\circl{dataflowcolor}{ } \circl{nodecolor}{ } Dataflow- or node-level explanations}
To support transparency, \systemname enables users to request natural language explanations for individual nodes or entire sections of the dataflow.
These are generated by prompting the LLM with the relevant parts of the specification and asking it to summarize the logic, purpose, and expected behavior of that component.
Users can also prompt for debugging ideas when facing actual or potential errors. These requests trigger focused prompts to the LLM, which in turn generates context-aware responses.
This aligns with \rectangle{black}{DG3-E}, ensuring that users can inspect and verify the AI-generated behavior of the system.
This is particularly useful when users are unfamiliar with certain operations or wish to validate nodes before proceeding, supporting \rectangle{black}{DG5-CC}.
%

\myparagraph{\circl{dataflowcolor}{ } \circl{nodecolor}{ } \circl{parametercolor}{ } Provenance and data inspection}
\systemname extends Curio's provenance model to capture versioned snapshots of the dataflow specification during key interactions.
Every time a user accepts an LLM suggestion, modifies a task or subtask, or changes a node's content, \systemname automatically creates a snapshot of the full specification, including nodes, edges, subtasks, and metadata.
These snapshots are then organized into a history tree that allows users to trace how their dataflow has evolved over time, supporting \rectangle{black}{DG4-E}. Users can browse earlier versions, inspect changes, and revert to previous versions if needed.
Each snapshot is timestamped and labeled with the triggering event or user prompt that led to the change.
\systemname also supports data-level inspection for each node. Based on the specification schema, it offers standard visualizations for possible outputs. Users can click a magnifying glass icon on a node to view its data, supporting \rectangle{black}{DG3-E}.
\subsection{The \systemname system}
\label{sec:system}

\systemname builds on Curio's dataflow framework~\cite{moreira2024curio}, adding alignment-driven features and a language-based interaction model.

\subsubsection{System architecture}

\systemname comprises a Python-based backend server and sandboxed execution server, with a React frontend.
\review{Most nodes use Python, while visualization nodes rely on Vega-Lite or UTK.}
\review{The backend manages dataflow logic, LLM interactions, data, provenance, and specification. Data remains in the backend in binary format, transferred internally between nodes, and moves to the frontend when a visualization node requests it.}
The sandbox server executes user or LLM-generated Python code in isolation.
The frontend provides an interactive interface for dataflow authoring, inspection, and LLM communication.

\subsubsection{LLM integration and prompting strategies}

\systemname relies on the LLM as a semantic engine for generating, modifying, and explaining parts of the dataflow. It supports multiple actions: task definition, subtask refinement, dataflow generation, debugging, and suggestions.
To make these actions reliable, the backend constructs each LLM request from three components: (1) A preamble, describing \systemname's internal logic, specification schema, UTK and Vega-Lite grammar structure, and system constraints; (2) A feature-specific prompt, tailored to the user's current action; (3) The input context, including the current dataflow specification, task and subtasks, expected input / output types, and node content.
The dataflow specification will be used to constrain the LLM responses.
To improve response quality, we use: (1) few-shot prompting with examples, (2) contextual priming with specs and metadata, (3) subprompt decomposition for multi-turn tasks, (4) role prompting to guide the LLM as an assistant, and (5) negative prompting to discourage vague or irrelevant outputs. The LLM can access dataset names and metadata, while persistent interaction is limited to referencing previous messages within the assistant chat.
%
%
%
Currently, \systemname uses OpenAI's gpt-4o-mini.

\begin{figure*}[t!]
    \centering
    \includegraphics[width=1\linewidth]{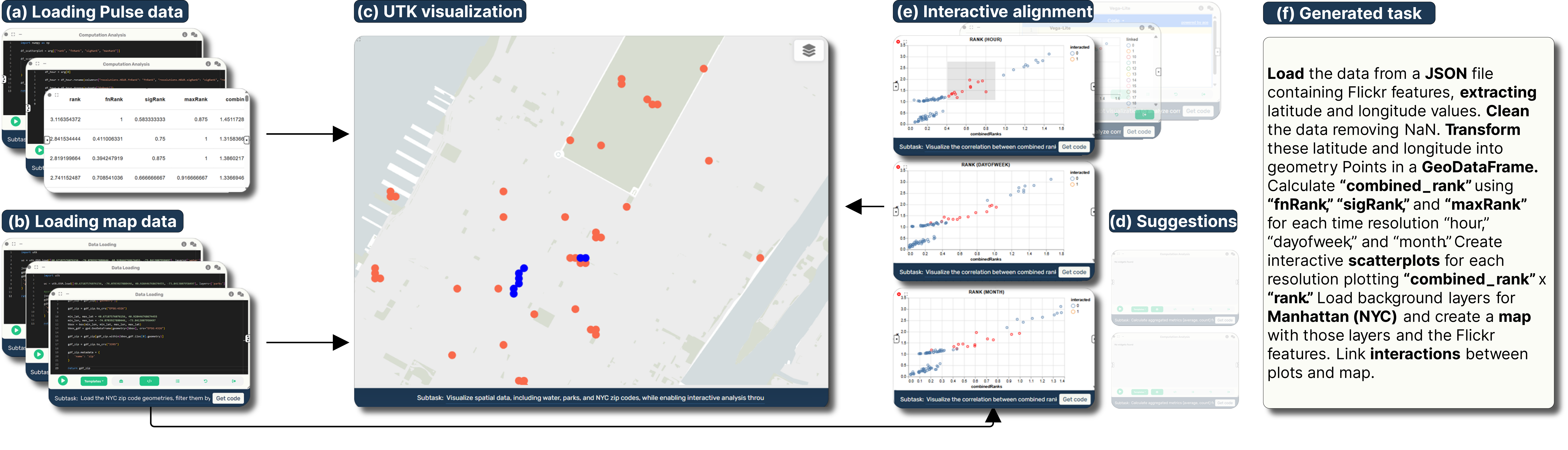}
    \vspace{-0.5cm}
    \caption{Using \systemname to reproduce Urban Pulse. (a) The user writes code to load and parse pulse data. (b) Background layers are added, and subtasks are generated using \textbf{task \& subtask definition}.  (c) A UTK node is added to visualize the data. (d) The user requests \textbf{connection suggestions} from the node and accepts a Vega-Lite recommendation. (e)  The Vega-Lite subtask is refined to include interactive selection, and the user uses \textbf{code generation} to implement it. (f) The task is continuously updated via \textbf{task \& subtask definition}, resulting in a final task.}
    \vspace{-0.5cm}
    \label{fig:case2}
\end{figure*}

\subsubsection{Frontend and interaction features}

The frontend interface presents a canvas-based authoring environment, integrated with LLM-powered and manual editing controls (Figure \ref{fig:features}). 

\myparagraph{Canvas area}
The canvas lets users build dataflows by dragging and connecting nodes. LLM-suggested nodes appear semi-transparent with an ``Accept Suggestion'' button. 
The node area provides node types and file controls for creating, uploading, and exporting dataflows.
A ``Dashboard Mode'' hides edges for presentation. Users define or edit tasks in natural language in the task editor, which highlights key terms and can be hovered to reveal their inferred connections. A ``Generate Suggestions'' prompts the LLM to scaffold the dataflow.
%
%
%
%
%

\myparagraph{Node area}
A node exposes interaction panels for specifying its purpose, input, and output types. Users can describe a subtask in natural language and prompt the LLM to generate node logic. A ``Suggest Connection'' button allows the LLM to propose next steps. An ``Explanation Tab'' provides an LLM-based summarization of the node. A magnifying glass lets users preview the output with a standard visualization.

%

\myparagraph{LLM assistant chat}
The right side of the interface hosts the LLM assistant chat, where the user can interact with the framework by revising tasks, clarifying suggestions, or exploring alternatives. This is the only LLM interaction that preserves chat history across turns.

\myparagraph{Provenance controls}
%
%
%
A version viewer lets users track dataflow provenance. LLM-triggered edits create snapshots of the specification, allowing users to inspect, compare, or revert changes (Figure~\ref{fig:teaser}(d)).

\section{Evaluation}



\subsection{Usage scenarios}


%
%
%
%
%
%
We present three real-world scenarios co-developed with urban experts who co-authored this paper.
%
We use inline indicators for interactions at different scopes: \circl{dataflowcolor}{ } dataflow, \circl{nodecolor}{ } node, and \circl{parametercolor}{ } parameter.
\review{Each scenario uses a different entry point with LLM augmentation: Scenario~1 starts with a chat task and dataflow request; Scenario~2 with sketching a dataflow from a paper diagram; and Scenario~3 with creating nodes to load data.}
%

\subsubsection{Scenario 1: Analyzing access with Project Sidewalk}
\label{sec:scenario-acessibility}

%
%
%
Crowdsourced platforms like Project Sidewalk~\cite{saha2019project} help fill data gaps by having volunteers annotate sidewalk features. However, such data introduces uncertainty, especially from contributor disagreements.
Using \systemname, we build a dataflow that combines multiple analytical and visualization layers -- such as comparing scores across neighborhoods and inspecting confidence levels.
In this scenario, the user begins by engaging with the LLM assistant in the chat-based interaction (Figure~\ref{fig:case1}(a)).
The user starts with a vague task: ``I want to load and visualize three datasets from Project Sidewalk with data observations as CSV, neighborhoods as GeoJSON, and streets as GeoJSON.''
The LLM uses metadata from uploaded files to ground the conversation and follows up with clarification questions to help the user structure a task \circl{dataflowcolor}{ }. Once satisfied (Figure~\ref{fig:case1}(b)), the user applies it to the dataflow \circl{dataflowcolor}{ }. The system proposes nodes for data loading, spatial joins, filtering, aggregation, and multi-view visualization (Figure \ref{fig:case1}(c,d,e)).
The user reviews and accepts most suggestions, editing subtasks to better reflect evolving goals (Figure \ref{fig:case1}(f), \circl{nodecolor}{ }). These edits automatically update the high-level task (Figure \ref{fig:case1}(h), \circl{dataflowcolor}{ }), keeping the specification aligned. For each node, the user prompts the LLM to generate code \circl{nodecolor}{ }, specifies expected input/output types \circl{parametercolor}{ }, and refines as needed.
During this process, the user writes a custom subtask for a new module \circl{nodecolor}{ } and receives a warning indicating that the logic could be modularized into separate operations -- splitting a transformation that filters and normalizes data into two nodes  (Figure~\ref{fig:case1}(e), \circl{nodecolor}{ }).
%
%
By the end, the dataflow supports dynamic comparison of accessibility and uncertainty. Tight task-subtask-dataflow alignment led to a clearer final task description, showing how iterative refinement improved human-AI alignment (Figure~\ref{fig:case1}(b,e)).

\subsubsection{Scenario 2: Topology exploration of social media data}





%
Urban Pulse~\cite{miranda2016urban} is a VA framework using topological techniques to identify and compare urban activity ``pulses.'' It models activity as a spatiotemporal scalar field, extracting peaks across time scales. The system includes coordinated views: a map of pulses, a scatter plot for comparison, and a line plot for trends.
\systemname enables users to recreate and extend such workflows by combining visual scaffolding, AI-assisted code generation, and synchronization between process-level actions and high-level task definitions. 
In this scenario, we illustrate how a user reproduces key elements of Urban Pulse’s workflow while leveraging \systemname's process alignment features to ensure that the resulting system remains coherent, modular, and explainable (Figure \ref{fig:case2}).
The user begins by uploading the Flickr dataset used in the original study. Unlike the previous scenario, they choose not to define a task up front via the LLM assistant. Instead, they take a bottom-up approach, sketching the initial dataflow structure (Figure \ref{fig:case2}(a,b), \circl{dataflowcolor}{ }) based on the original paper’s diagrams. 
For simple preprocessing steps, such as loading and formatting the data, the user writes Python code into node editors and executes them (Figure \ref{fig:case2}(a,b), \circl{nodecolor}{ }). 
As each node is run, the LLM analyzes the node content and automatically generates a subtask (Figure \ref{fig:case2}(a,b), \circl{nodecolor}{ }), incrementally constructing a task description in the background (Figure \ref{fig:case2}(f), \circl{dataflowcolor}{ }).
Throughout, the user frequently uses the connection suggestion to explore next steps (Figure \ref{fig:case2}(e), \circl{dataflowcolor}{ }).  These LLM-generated prompts help expand the dataflow based on the task and dataflow (Figure \ref{fig:case2}(d), \circl{dataflowcolor}{ }). The user accepts or edits suggestions and updates subtasks \circl{nodecolor}{ } as needed, resulting in three dataflow branches for Urban Pulse's time resolutions: hour, day, and month (Figure \ref{fig:case2}(d), \circl{dataflowcolor}{ }).
The LLM recommends visualizing each branch with a scatter plot (Figure \ref{fig:case2}(d), \circl{nodecolor}{ }). 
After accepting this suggestion, the user modifies the subtask to include interactive selections. The final output mirrors the Urban Pulse system, supporting multi-scale activity exploration.

To build the map-based view, the user adds nodes for loading background layers (Figure \ref{fig:case2}(b), \circl{nodecolor}{ }), merges them with pulse point data \circl{nodecolor}{ }, and visualizes the result using a UTK node (Figure \ref{fig:case2}(c), \circl{nodecolor}{ }), with parameters \review{exposed} to the user \circl{parametercolor}{ }. They complete the workflow by creating interaction edges between the scatter plots and the map view \circl{dataflowcolor}{ }, allowing selections in one view to update the others, turning what would normally be a technically demanding coordination task into a visual connection.
Even though the user never explicitly defined a task in advance, the task and subtasks are continually synchronized (Figure \ref{fig:case2}(f), \circl{dataflowcolor}{ }\circl{nodecolor}{ }), preserving intent and clarity across modules.
This scenario demonstrates \review{Urbanite’s} support for process-first workflows, where a user incrementally constructs a system and achieves strong alignment through bottom-up iteration and AI collaboration (Figure \ref{fig:case2}).

\subsubsection{Scenario 3: Impacted houses in flooding simulation}


%
Urban VA is crucial for environmental risk assessment~\cite{de_souza_prowis_2022, boorboor_submerse_2024}. Climate experts use models to simulate real-world phenomena, and libraries like SynxFlow~\cite{synxflow} simplify flooding simulations. 
However, non-experts still struggle with terminology and data, while domain experts need better ways to explore and visualize results. We show how a user can use \systemname's dataflows, UTK visualizations, and LLM-powered features to interpret simulations and build complex visualizations.
%
%
%

The user starts by creating nodes to load data for a small village in the Chicago area \circl{dataflowcolor}{ }. The data is used in a computation node (Figure \ref{fig:teaser}(b), \circl{nodecolor}{ }) to run a SynxFlow flooding simulation for an extreme storm event.
As nodes execute, the LLM generates subtask descriptions \circl{nodecolor}{ } which are automatically incorporated into the task, maintaining the broader context (Figure \ref{fig:teaser}(f), \circl{dataflowcolor}{ }). 
An additional node (Figure \ref{fig:teaser}(a), \circl{nodecolor}{ }) fetches 3D building data from OpenStreetMap using the UTK API. To check the result, the user uses the data inspection tool (Figure \ref{fig:teaser}(a), \circl{parametercolor}{ }).
Finally, two UTK visualizations (Figure \ref{fig:teaser}(c), \circl{nodecolor}{ }) are created: one providing an overview of the simulation and another highlighting flood risk for individual houses. 
Wanting to change key simulation parameters, the user \review{goes} back to a previous dataflow version using provenance (Figure \ref{fig:teaser}(d),  \circl{dataflowcolor}{ }), updating the task description (Figure \ref{fig:teaser}(f), \circl{dataflowcolor}{ }) and consequently the subtasks of respective nodes \circl{nodecolor}{ }.
Based on the new subtasks, the user requests updated code for the simulation nodes \circl{nodecolor}{ } from the LLM. The provenance tree now has two branches (Figure \ref{fig:teaser}(d), \circl{dataflowcolor}{ }), allowing easy comparison of simulation configurations.
To share this workflow with collaborators, the user generates node-level explanations (Figure \ref{fig:teaser}(b), \circl{nodecolor}{ }) for simulation nodes and a dataflow-level explanation (Figure \ref{fig:teaser}(e), \circl{dataflowcolor}{ }).
Finally, the dataflow is exported, and markdown files with explanations are generated.
In the end, the user creates a dataflow to simulate floods in the Chicago area, with an overview visualization and a 3D view of affected buildings. This scenario demonstrates \systemname's ability to enhance evaluation alignment and collaboration through provenance tracking and automatic documentation (Figure~\ref{fig:teaser}).




\subsection{Experts' feedback}

To evaluate \systemname, we conducted semi-structured interviews with six urban experts using one usage scenario, gathering feedback on each feature. 
%
%
None were paper authors; all had programming experience and had used LLMs.
%
%
%
Participants included three civil engineering PhD students \expert{E1-3}, an urban planning faculty member \expert{E4}, a water resource engineering researcher \expert{E5}, and a computer scientist specializing in urban accessibility \expert{E6}.
Overall, they expressed positive impressions towards \systemname, appreciating its ease of use and accessibility.
Regarding the \textbf{dataflow generation}, \expert{E1} positively remarked, ``It's really amazing because usually we have to spend a long time defining these things by ourselves,'' highlighting that the framework ``gives us an initial version we can work around.'' 
Similarly, \expert{E2} noted the novelty of the \review{\textbf{dataflow generation}} feature, emphasizing, ``this doesn't exist in tools that we have.'' \expert{E4} appreciated the conversational nature of the dataflow generation, mentioning that ``being prompted on figuring out more detailed tasks is helpful,'' since in her experience, ``LLMs sometimes do something that has nothing to do with what you're asking.''

Concerning the \textbf{task and subtask definition}, \expert{E6} found the suggestion of subtasks useful, observing, ``The idea of suggesting subtasks is interesting, and the user will even learn through them by visualizing the isolated subtasks.'' \expert{E4} also praised the concept, emphasizing the benefit of syncing code and task, ``that's where information gets lost in a lot of projects (...) you update the code but you don't update the notes, so keeping them synced is very helpful.'' 
\expert{E2} aligned positively with the subtasks feature, as it closely matches his workflow of ``breaking down a problem into distinct parts, visually plotting it, seeing what needs to be done next.''
\expert{E5} mentioned that this feature ``simplifies the task'' of creating dataflows, stressing that ``from a project perspective, this immediately cleans up how we understand our dataflow.''

On \textbf{code generation}, \expert{E4} underscored its practical value, appreciating ``the simplicity of having LLM helping me with the task.'' Similarly, \expert{E3} found significant value in context integration, stating, ``Usually you have your VS Code, and then you go to ChatGPT, but I appreciate that the code generation happens inside the tool and in the context of a node.'' \expert{E4} particularly appreciated the structured coding environment provided by the tool: ``Usually I use individual scripts for coding, \systemname structures them clearly within the same canvas.''
With respect to the \textbf{connection suggestions}, \expert{E1} acknowledged how this feature enables incremental refinement of workflows. \expert{E3} further highlighted the creative potential of this capability, calling it ``a very interesting feature for brainstorming and exploring analytical possibilities.''
%
%
\expert{E4} especially valued the \textbf{explanation} and \textbf{provenance} features for supporting reproducibility and transparency: ``My mind is going to things like reproducibility and open science, linking code with clear descriptions of what it does.''
She also highlighted the clarity: ``It contextualizes the subtasks within the big dataflow,'' the ease of sharing: ``You can easily share with others,'' and support for iteration: ``I can go back and iterate on specific aspects easily.''
%
%
%
The urban experts also expressed certain reservations. For example, \expert{E5} mentioned that technical terms differ across countries, so that terminology might not be factored in by LLMs.
Both \expert{E3} and \expert{E5} mentioned concerns regarding the short-term caching of LLM context, with \expert{E5} mentioning that ``when the short-term cache ends, if the user returns later, the context might already be lost.''

\subsection{Quantitative evaluation}

\review{To assess Urbanite's effectiveness in translating high-level user intent into executable visual analytics workflows, we conducted an evaluation focusing on the core stages of LLM-powered generation: task semantic alignment, subtask coverage, and dataflow quality. In order to achieve this, we first selected ten representative urban VA papers. From each of these papers, we manually extracted the core user intent that the corresponding VA system aimed to support. Using Urbanite, we then: (1) generated a task description based on this user intent, (2) decomposed it into subtasks, and (3) matched these subtasks to a dataflow graph.}

\myparagraph{Methodology}
\review{We recruited five experienced Computer Science researchers as evaluators; none are authors of this paper: three faculty and two PhD students. They were instructed to assess the generated outputs and how they translated intent into a dataflow.}
\review{Each paper was assigned to two evaluators, who scored three aspects on a 3-point ordinal scale, where higher scores indicate better quality: \textbf{Semantic alignment} measured how well the generated task description reflected the original user intent, ranging from ``Misaligned'' (0) to ``Fully aligned'' (2). \textbf{Coverage of subtasks} assessed whether the subtask decomposition included all essential steps for task completion, from ``Incomplete'' (0) to ``Complete'' (2). \textbf{Flow quality} evaluated whether the dataflow's structure supported task execution, ranging from ``Invalid flow'' (0) to ``Functional and coherent'' (2). Additionally, evaluators could flag outputs for hallucination if fabricated elements were present.}

\myparagraph{Results \& takeaways}
\review{\systemname achieved strong alignment with user intent across the evaluated cases, with an average \textbf{semantic alignment score of 1.65 (SD=0.32)} on the 0--2 scale. In most cases, the generated task descriptions captured the essence of the original intent effectively.
For \textbf{subtask coverage}, \systemname achieved an average score of \textbf{1.6 (SD=0.37)}, suggesting that the system was generally able to decompose user intent into comprehensive subtasks, although a few cases exhibited partial coverage where additional subtasks would enhance completeness.
The \textbf{flow quality} evaluation had an average of \textbf{1.5 (SD=0.45)}, indicating that the generated dataflows were typically functional and coherent, but with some instances requiring refinement.
Across all evaluations, we identified three noteworthy mismatches between the extracted tasks and the original user intent:
For Ferreira et al.~\cite{ferreira2013visual}, the extracted task incorrectly inferred that the analysis was conducted using hourly or daily patterns, which was not specified in the original intent.
For Ferreira et al.~\cite{ferreira2015urbane}, the extracted task incorrectly assumed that sky exposure data was available through an open data portal and omitted a necessary computation step.
Finally, for Konev et al.~\cite{konev2014run}, the extracted task failed to capture a critical aspect of the user intent: the need to run and control simulations. Additionally, the generated dataflow was overly vague and lacked sufficient detail to support execution.
Full results are in the supplementary material.
}

\section{Conclusions}

\myparagraph{Reflection on design goals}
The design goals of \systemname were defined to bridge the gap between human intent and AI behavior across all stages of an urban VA dataflow.
%
%
Regarding \rectangle{black}{DG1-S}, by allowing users to articulate analytical intent in natural language, \systemname lowers the entry barrier to the creation of these dataflows. \systemname does not require users to translate ideas into technical components from the outset. Instead, the framework shifts the burden of interpretation to the AI. Despite this automation, the framework still gives users control over the authoring process, as highlighted by \rectangle{black}{DG2-P}; \systemname offers suggestions rather than decisions, so all AI-generated content is inspectable and editable.
As highlighted in our experts' feedback, transparency is key in fostering trust in AI-assisted workflows.
\systemname directly addresses \rectangle{black}{DG3-E}; whether through explanations or data inspections, users are given visibility into what the dataflow is doing.
To support trust, transparency, and reproducibility, \systemname incorporates provenance of dataflow specifications, which aligns with \rectangle{black}{DG4-E}. By tracing both data transformations and human-AI interactions over time, the framework allows users to compare past states, revert changes, and understand the evolution of their dataflows.
%
%
As noted in \rectangle{black}{DG5-CC}, \systemname supports rephrasing tasks and revising AI-generated dataflows, enabling alignment as an iterative dialogue where human and AI gradually converge.
For alignment stages in \rectangle{black}{DG6-CC}, \systemname supports multi-level interaction by keeping scopes consistent. Users can start with a broad task, refine subtasks, switch between dataflow and node logic, and explore results at different levels. Changes in one scope automatically update others to ensure consistency.
%
%
\review{While our evaluation revealed that \systemname occasionally made mistakes (e.g., inferring unavailable datasets, omitting specific steps, overly broad flows), these instances were limited. These results suggest that \systemname can effectively translate high-level user intent into executable workflows, providing a strong starting point for urban VA. Importantly, such imperfections highlight the value of \emph{\systemname's in-the-loop, iterative refinement process}, where users can easily identify and adjust AI-generated outputs. Rather than seeking a perfect one-shot generation, \systemname is designed to facilitate progressive alignment between human intent and system behavior.}

\myparagraph{Limitations}
\review{While \systemname presents a novel approach to urban VA, we acknowledge a few limitations.
The visualizations generated are primarily constrained by the capabilities of Vega-Lite and UTK. This can make it challenging to create more complex, custom-layered visualizations that require very specific visual encodings. Additionally, \systemname's performance is primarily constrained by the underlying libraries utilized within each dataflow node. We also note that the framework does not yet support the automatic positioning or reorganization of dataflow nodes through LLM interaction, which could pose a visual burden for very comprehensive dataflows.
}


\myparagraph{Future work}
\review{Building upon \systemname's foundational capabilities, we envision several avenues for future work. We are particularly interested in exploring intelligent dataflow layout, alongside a detailed investigation into the cognitive burden of dataflow interaction, especially for complex analytical workflows. Another compelling research avenue lies in a comparative study of different interaction modalities, including dataflow-centric approaches, notebook-based systems, grammar-based interfaces, and our LLM-driven natural language interaction. In addition, investigating these avenues with a larger cohort of users would be valuable for expanding upon our findings and ensuring broader applicability across different expertise levels.}

\section*{Acknowledgments}
We thank the reviewers for their constructive feedback.
This work was supported by the U.S. National Science Foundation (Awards \#2320261, \#2330565, \#2411223) and conducted as part of the Open-Source Cyberinfrastructure for Urban Computing (OSCUR) project.




\bibliographystyle{abbrv-doi-hyperref}

\bibliography{references}

\end{document}